\documentclass[photonics,article,accept,pdftex,moreauthors]{Definitions/mdpi} 
\firstpage{1} 
\pubvolume{1}
\issuenum{1}
\articlenumber{0}
\pubyear{2026}
\copyrightyear{2026}
\datereceived{18 March 2026} 
\daterevised{23 April 2026} % Comment out if no revised date
\dateaccepted{28 April 2026} 
\datepublished{ } 
\Title{Transit Noise in Spin Squeezing Experiments with Coated Rubidium Vapor Cell} %Attention: Title altered

\Author{Yujie %MDPI: Please carefully check the accuracy of names and affiliations. Please notice that any changes about authorship, affiliations, and corresponding authors are not allowed after the paper has been accepted, including adding, removing, or changing the order.
 Ji $^{1,2}$, Peiying Li $^{1,2}$, Yanhong Xiao $^{1,2,3}$%MDPI: The numbers associated with the authors’ affiliations should appear in numerical order. Please check the suggested changes. Same as below.
, Yuzhuo Wang $^{4,}$* and Junlei Duan $^{1,2,3,}$*}

\AuthorNames{Yujie Ji, Peiying Li, Yanhong Xiao, Yuzhuo Wang and Junlei Duan}

\address{%
$^{1}$ \quad State Key Laboratory of Quantum Optics Technologies and Devices, Institute of Laser Spectroscopy, \linebreak Shanxi University, Taiyuan 030006, China; jiyujie@sxu.edu.cn (Y.J.); lipeiying@sxu.edu.cn (P.L.); yxiao@fudan.edu.cn (Y.X.)\\
$^{2}$ \quad Collaborative Innovation Center of Extreme Optics, Shanxi University, Taiyuan 030006, China\\
$^{3}$ \quad State Key Laboratory of Surface Physics, Department of Physics, Fudan University, Shanghai 200438, China\\
$^{4}$ \quad School of Physics and Electronic Engineering, Sichuan University of Science \& Engineering, \linebreak Zigong 643000, China}

\corres{Correspondence: wangyz@suse.edu.cn (Y.W.); jlduan@fudan.edu.cn (J.D.)}

\abstract{Spin squeezing can suppress quantum projection noise via interparticle entanglement, therefore enabling measurement sensitivities %AUTHOr: We have changed 'achieving sensitivity' to 'enabling measurement sensitivities'.
beyond the standard quantum limit. In practice, however, the Gaussian and finite intensity profiles of the optical %AUTHOr: We have added 'optical'.
probe beam induce spatially inhomogeneous atom-light %AUTHOR: All instances of 'atom–light' in the manuscript have been unified with a hyphen.
interactions. As polarized atoms move within a vapor cell, %AUTHOr: We have changed 'within a coated rubidium vapor cell' to 'within a vapor cell'.
they experience position-dependent optical intensities, generating transit noise that limits spin squeezing performance. Here, we investigate the transit noise %AUTHOr: We have removed 'of atoms'.
in a coated rubidium vapor cell through combined theoretical analysis and experimental measurements. By varying the probe beam diameter, we quantify the dependence of transit noise on beam size and atomic Larmor frequency. Our results show that, for a vapor cell with fixed dimensions, the transit noise increases as the probe beam spot area decreases. Moreover, when the Larmor frequency is below the characteristic linewidth of the transit noise, the noise contribution becomes larger. We further calculated and measured spin squeezing for different beam sizes and found an experimental difference of $2.7 \pm 0.2$ dB between 2~mm and 0.6~mm, similar %AUTHOr: We have changed 'compared' to 'similar'.
to the theoretical prediction of $3.0 \pm 0.3$ dB. Theoretical analysis under conditions of stronger squeezing shows that transit noise becomes an even more critical limiting factor. These results provide practical guidance for optimizing probe beam parameters and suppressing transit noise in spin squeezing experiments.}

%AUTHOr: We have removed 'probe'.

\begin{document}

%%%%%%%%%%%%%%%%%%%%%%%%%%%%%%%%%%%%%%%%%%
\section{Introduction}

Precision measurement is important for both fundamental physics research~\citep{bu1,bu2} and practical applications in external field sensing~\citep{bu3,bu4}. Atoms have received much attention in the field of precision measurement due to their advantages, such as simple energy level structures, narrow transition linewidths, sensitivity to various external fields, and ease of state manipulation and readout~\citep{ref1,ref2,ref3,ref4}. The core advantage of quantum precision measurement is the possibility to harness %AUTHOr: We have changed 'its ability to utilize' to 'the possibility to harness'.
quantum properties like coherence and entanglement~\citep{ref5} to surpass the precision limit of classical measurement techniques---the standard quantum limit (SQL). Among these, spin squeezing~\citep{ref6,ref7}, as a technique for quantum entanglement, can utilize correlations between particles~\citep{ref8} to suppress noise, thereby breaking the SQL. In recent years, the field of spin squeezing has witnessed %AUTHOr: We have changed 'spin squeezing has achieved' to 'the field of spin squeezing has witnessed'.
significant progress in applications such as atomic clocks~\citep{ref9,ref10}, atom interferometers~\citep{ref11}, and atomic magnetometers~\citep{ref12,ref13}.

Spin-squeezed states have been extensively investigated in a wide range of physical platforms~\citep{ref14,ref15,ref16,ref17,ref18,ref19,ref30}. Hot atomic ensembles are particularly attractive for precision measurements owing to their large atom numbers. However, in thermal vapors, the rapid atomic motion causes atoms to repeatedly enter and exit the probe beam region, giving rise to transit noise. Moreover, the spatially varying intensity of the probe beam leads to temporally and spatially inhomogeneous atom-light interactions as atoms move across the beam profile. Consequently, the realization of spin squeezing in such systems requires effective suppression of transit noise and interaction inhomogeneity.
In 2020, Tang et~al.~\citep{ref20} employed a variable-diameter flat-top probe beam to investigate atomic transit noise in an unpolarized vapor in a coated cell. %AUTHOr: We have added 'in a coated cell'.
They demonstrated that the total spin noise power scales inversely with the beam area, whereas the Ramsey peak power remains independent of it. In 2023, Wang et~al.~\citep{ref21} utilized a similar flat-top configuration to precisely control the effective atom number in an atomic superheterodyne microwave receiver, %AUTHOr: We have added 'microwave'.
revealing that transit noise constitutes a fundamental limitation to its sensitivity. In 2024, Yang et~al.~\citep{ref22} studied transit noise in a Rydberg receiver and showed that its spectral bandwidth scales inversely with the beam waist, while the noise amplitude depends on the probe intensity. In the same year, B\ae{}rentsen et~al.~\citep{ref23} compared the flat-top and Gaussian probing of a Cs ensemble and identified spatially uniform illumination as an effective approach for suppressing transit noise. Despite these advances, the role of transit noise in spin squeezing experiments, to the best of our knowledge, has not been studied. 

In this work, we investigated the influence of atomic transit noise on spin squeezing with polarized atoms by varying the diameter of a Gaussian probe beam, combining theoretical analysis with experimental measurements. Although the qualitative trend of larger transit noise for smaller probe beams is physically intuitive, its effect on spin squeezing is not determined solely by the transit noise background. In QND-based spin squeezing, the atomic transit process also modifies the temporal correlation of the detected optical signal and thereby affects the conditional variance reduction. In particular, we found that as the laser beam size varies, the change in the degree of spin squeezing in decibels (dB) is larger than the corresponding change in the transit noise level. First, we measured the atomic spin noise spectra at different Larmor frequencies and beam sizes, and the results show that the transit noise increases for smaller beam diameters and becomes more pronounced at lower Larmor frequencies. Then, we evaluated the achievable spin squeezing at a Larmor frequency of 500~kHz, observing an experimental difference of $2.7 \pm 0.2$ dB %AUTHOR: We have added the unit “dB” in the relevant place.
between probe beam diameters of 0.6~mm and 2~mm, similar to the theoretical value of $3.0 \pm 0.3$ dB, both far exceeding the corresponding changes in the level of transit noise (less than 1~dB). These findings demonstrate that transit noise constitutes a significant limitation to spin squeezing and that the probe beam size plays a crucial role in determining the maximum achievable squeezing.

%%%%%%%%%%%%%%%%%%%%%%%%%%%%%%%%%%%%%%%%%%
\section{Theory}
\subsection{The Input-Output Relation of a Quantum Nondemolition Measurement}
The coated $^{87}$Rb vapor cell used in our experiment contains %AUTHOR: We have corrected a spelling error.
about $ 10^{10}$ atoms. In such a large atomic ensemble, the atoms are individually indistinguishable, and the atomic state of the entire system can be described by a collective spin operator. The collective angular momentum operator of the atomic ensemble is defined as: %MDPI: 1. Please recheck all equations and make sure there are no duplicated equations in the whole manuscript. 2. Please carefully check variable formatting (italic, bold, subscript, uppercase, etc.) throughout the manuscript to ensure the formatting is consistent and revise if needed. Thanks.
%AUTHOR: 1. We have rechecked all equations throughout the manuscript and confirmed that there are no duplicated equations. 2. We have carefully reviewed the formatting of all variables throughout the manuscript. Inconsistencies have been corrected, and the formatting is now consistent. The changes have been highlighted in yellow.

\begin{equation}
\hat{\mathbf{J}}_n=\sum_{i=1}^{N_A}\hat{\mathbf{j}}_n^{(i)},
\label{eq:Jdef}
\end{equation}
where $n=x,y,z$, $\hat{\mathbf{j}}_n^{(i)}$ is the dimensionless spin operator %AUTHOR: We have changed the description of $\hat{\mathbf{j}}_n^{(i)}$ from "angular momentum" to "dimensionless spin operator".
of the $i$-th atom, and $N_A$ is the total number of atoms in the vapor cell.

We assume %AUTHOR: We have changed 'Suppose' to 'We assume'.
that the atoms are optically pumped into the stretched state of the \linebreak $F=2$ hyperfine manifold and are highly polarized along the $x$-direction. In this case, the angular momentum component $\hat{J}_x$ is treated as the classical component, $\langle \hat{J}_x\rangle = 2N_A$; the orthogonal components $\hat{J}_y$ and $\hat{J}_z$ exhibit quantum fluctuations, satisfying $\langle \hat{J}_y \rangle = \langle \hat{J}_z \rangle = 0$ and $\operatorname{Var}(\hat{J}_y) = \operatorname{Var}(\hat{J}_z) = |\langle\hat{J}_x\rangle|/2$~\citep{bu6} 
%AUTHOR: We have added absolute value signs around $\langle \hat{J}_x \rangle$.
and are called the quantum components since the angular momentum in the $x$-direction is much larger than those in the $y$- and $z$-directions. %AUTHOR: We have removed the sentence "Here and in the following, we set $\hbar = 1$".
We describe the atomic ensemble in this state as a coherent spin state (CSS). For such a highly polarized atomic ensemble, according to the Holstein–Primakoff (HP) approximation~\citep{ref24}, we can define the %AUTHOR: We have added 'the'.
two quantum components as a pair of canonically conjugate operators:
\begin{equation}
\hat{x}_A=\frac{\hat{J}_y}{\sqrt{|\langle J_x \rangle|}},\qquad
\hat{p}_A=\frac{\hat{J}_z}{\sqrt{|\langle J_x \rangle|}},\qquad
[\hat{x}_A,\hat{p}_A]=\mathrm{i} .
%AUTHOR: We have added absolute value signs around $\langle \hat{J}_x \rangle$ in the denominator of $\hat{x}_A$ and $\hat{p}_A$.
\label{eq:HP}
\end{equation}
%AUTHOR: Here and in the following, we set $\hbar = 1$.

For the quantum %MDPI: Please check if the no indent format should be retained after the equation, please check the whole text.
%AUTHOR: Thank you for your comment. We have decided not to retain the "no indent" format after equations and have changed it to an indented format.
 state of light, we employ the Stokes operators to describe the polarization state. For a light beam propagating along the \(z\)-direction, the Stokes operators can be defined as:
\begin{equation}
\hat{S}_x = \frac{1}{2} (\hat{n}_x - \hat{n}_y), \quad
\hat{S}_y = \frac{1}{2} (\hat{n}_{+45^\circ} - \hat{n}_{-45^\circ}), \quad
\hat{S}_z = \frac{1}{2} (\hat{n}_{\text{rh}} - \hat{n}_{\text{lh}}).
\label{eq:Stokes}
\end{equation}
where $\hat{n}_x$ and $\hat{n}_y$ denote the photon numbers in the $x$ and $y$ linear polarizations, $\hat{n}_{+45^\circ}$ and $\hat{n}_{-45^\circ}$ denote the photon numbers in the $+45^\circ$ and $-45^\circ$ linear polarizations, $\hat{n}_{lh}$ and $\hat{n}_{rh}$ denote the photon numbers in the left and right circular polarization modes. As can be seen, \(\hat{S}_i\) is a dimensionless quantity. We can also define the Stokes vector per unit time, \(\hat{S}_i(t)\), whose dimension is \(1/\text{time}\), satisfying \(\hat{S}_i = \int_0^T \hat{S}_i(t) \, dt\), where \(T\) is the total duration of the probe light. Similar to the collective spin operator of atoms, the Stokes operators of light can also define a pair of canonically conjugate operators:
\begin{equation} 
\hat{x}_L(t)= \frac{\hat{S}_y(t)}{\sqrt{|\langle S_x(t)\rangle|}},\qquad
\hat{p}_L(t)=\frac{\hat{S}_z(t)}{\sqrt{|\langle S_x(t)\rangle|}},\qquad
[\hat{x}_L(t),\hat{p}_L(t')]= i\,\delta(t-t').
%AUTHOR: We have added absolute value signs around $\langle \hat{S}_x(t) \rangle$ in the denominator of $\hat{x}_L(t)$ and $\hat{p}_L(t)$.
\label{eq:LightQuad}
\end{equation}

In our system, the effective Hamiltonian for the atom-light interaction~\citep{ref25} can be written (schematically) as: %AUTHOR: We have removed the phrase "a sum of scalar, vector, and tensor interactions".
\begin{equation}
\hat{H}=\hat{H}^{(0)}+\hat{H}^{(1)}+\hat{H}^{(2)},
\label{eq:Hfull}
\end{equation}
where $\hat{H}^{(0)}$ is the scalar term (AC Stark shift), $\hat{H}^{(1)}$ is the vector term %We have removed 'Faraday'.
and $\hat{H}^{(2)}$ is the tensor term. In our theory and experiments, we consider only the vector interaction term, which represents the Faraday interaction between light and atoms, enabling the exchange of information between light and atoms~\citep{ref6,bu6}. The scalar term is included in an effective frequency shift. For the detuning used in the experiment, the tensor contribution was much smaller than the vector contribution and is neglected in the following analysis. In the far-detuned case, the Hamiltonian can be approximated as~\citep{bu12,ref25,ref18}:
\begin{equation}
\hat{H}_{\text{int}} = \hbar c a \int_0^L \hat{S}_z(z,t) \hat{j}_z(z,t) \, \rho A \, dz,
\label{eq:FaradayH}
\end{equation}
where $ a = -\frac{\Gamma \lambda^2}{8A\Delta \, 2\pi} a_1$, 
\(\hbar\) is the reduced Planck constant, \(c\) is the speed of light, \(\Gamma\) is the nature width of the excited state, %EE: Please check intended meaning has been retained in the prior sentence.
%AUTHOR: We have revised the definition of \(\Gamma\).
\(\lambda\) is the wavelength, \(A\) is the area of the atom-light interaction, \(\Delta\) is the detuning of the probe, \(a_1\) is the vector interaction coefficient, \(\hat{S}_z(z,t)\) is the Stokes operator density of the propagating light field with units of inverse length %AUTHOR: A explanation has been added for the symbol \(\hat{S}_z(z,t)\).
and \(\rho\) is the number density. According to the Heisenberg equation of motion and the propagation equation for Stokes operators, and after replacing the atomic collective spin operators and Stokes operators with their respective canonical quadrature operators, the differential equation for the atomic spin operator can be expressed as:
\begin{equation}
\begin{aligned}
\frac{d}{dt} \hat{x}_A(t) &= \kappa \cdot \hat{p}_L^{\text{in}}(t), \\
\frac{d}{dt} \hat{p}_A(t) &= 0, 
\end{aligned}
\label{eq:IO1}
\end{equation}
and the input-output relation for the atom-light interaction %AUTHOR: We have deleted a hyphen (e.g., changed "atom–light- interaction" to "atom–light interaction").
is obtained~\citep{ref26}:
\begin{equation}
\begin{aligned}
\hat{x}_L^{\text{out}}(t) &= \hat{x}_L^{\text{in}}(t) + \kappa \cdot \hat{p}_A(t), \\
\hat{p}_L^{\text{out}}(t) &= \hat{p}_L^{\text{in}}(t),
\end{aligned}
\label{eq:IO2}
\end{equation}
where \(\kappa = a\sqrt{|S_x(t)\cdot J_x(t)|}\) is the effective atom-light coupling strength in the idealized homogeneous coupling limit. A more detailed expression for \(\kappa\), showing its relationship to the optical power, is presented in Appendix~\ref{kappa and p}. More generally, however, atom-light coupling may vary in time and is then denoted by \(\kappa(t)\), which represents the instantaneous atom-light coupling strength. This time dependence can arise, for example, from temporal modulation of the probe or from the fact that atoms experience different local probe intensities at different times. Its explicit dependence on the atomic motion will be introduced in the following subsection. The effective coupling \(\kappa\) may be regarded as the spatially and temporally averaged value of the instantaneous coupling \(\kappa(t)\).

From the input-output relation, it can be seen that \(\hat{p}_A\) remains constant during the interaction process, so the interaction is non-demolishing with respect to \(\hat{p}_A\). Meanwhile, \(\hat{x}_L\) acquires information about \(\hat{p}_A\) through the interaction. Therefore, by measuring \(\hat{x}_L\), one can achieve a quantum nondemolition~\citep{bu8,bu9,bu10} (QND) measurement of \(\hat{p}_A\). The above derivation neglects the presence of an external magnetic field, so the Larmor frequency is zero. %AUTHOR: We have removed 'of atomic motion', 'at' and 'frequency'. 
By applying an external magnetic field, the measurement %AUTHOR: We have changed 'experimentally measured' to 'measurement'.
frequency can be shifted away from zero. Therefore, after adding a bias magnetic field along the $x$ direction, the Hamiltonian can be expressed as:
\begin{equation}
\begin{aligned}
\hat{H}_B = \hbar \Omega \hat{J}_x,
\end{aligned}
\label{eq:b}
\end{equation}
where $\Omega$ is the Larmor frequency. In the presence of a bias magnetic field, the atomic quadratures precess at the Larmor frequency $\Omega$, so that the measured collective spin component becomes a rotating quadrature:
\begin{equation}
\begin{aligned}
\hat{x}_{\text{L}}^{\text{out}}(t) &= \hat{x}_{\text{L}}^{\text{in}}(t) + \kappa(t)\left( \hat{p}_A(t) \cos(\Omega t) - \hat{x}_A(t) \sin(\Omega t) \right), \\
\hat{p}_{\text{L}}^{\text{out}}(t) &= \hat{p}_{\text{L}}^{\text{in}}(t).
\end{aligned}
\label{eq:b2}
\end{equation}

\subsection{Theoretical Model}

\begin{figure}[H]
%\centering
 \includegraphics[width=0.5\textwidth]{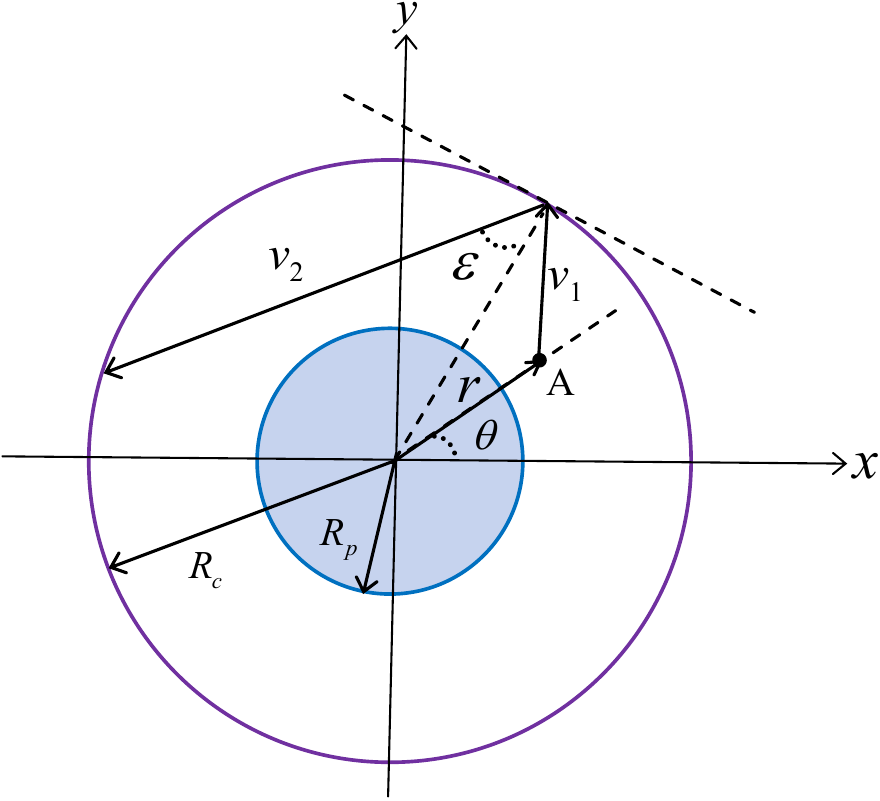}
\caption{Schematic diagram %MDPI: Please check if dashed line and blue area need add explanations.
%AUTHOR: We have checked the dashed line and blue area in the figure. Explanations have been added in the figure caption as needed. The changes are highlighted in yellow.
 of atomic motion inside a rubidium vapor cell. The blue area is the region covered by the light spot; the dashed lines are the tangent and normal lines, respectively.}
\label{fig:mon}
\end{figure}

We employ the Monte Carlo method to simulate the motion of atoms in the coated cell, tracking random motion and wall collisions. For a Gaussian probe beam propagating along the $z$-axis, since the cell length is much shorter than the Rayleigh length, we assume that the intensity is uniform along $z$ and varies mainly in the transverse plane.
Based on the simplified model, the schematic diagram of atomic motion is shown in Figure~\ref{fig:mon}. The cell has an effective cross-sectional radius \(R_c\), and the probe beam with radius \(R_p\) is centered within the cell. We assume that at the initial time, the atoms are uniformly distributed throughout the volume of the cell. The initial position of the atom is the point \(A\), which is described using polar coordinates \((r, \theta)\), where the square of the radial coordinate, $r^2$, is uniformly distributed in the range $[0, R_c^2]$, and the polar angle $\theta$ is uniformly distributed in the range $[0, 2\pi)$. The speed $v_1$ of the atom follows a two-dimensional Maxwell-Boltzmann~distribution:
\begin{equation}
f(v)=\frac{v}{\sigma^2}\exp\!\left(-\frac{v^2}{2\sigma^2}\right),
\label{eq:MB2D}
\end{equation}
where $\sigma^2 = k_B T/m$, $k_B$ is the Boltzmann constant, $m$ is the mass of a $^{87}$Rb atom, and $T$ is the temperature of the vapor.

\vspace{-5pt}

We further consider the description of atomic collisions within the coated cell. During their high-speed motion, atoms experience both atom-atom and atom-wall collisions. Given that for our operating temperature %AUTHOR: We have add 'for our operating temperature'.
the mean free path of the atoms is significantly larger than the dimensions of the vapor cell, the rate of atom-atom collisions is negligible compared to that of atom-wall collisions. Therefore, we omit the former to simplify the model. The collision process between atoms and the cell wall is simulated with a diffuse reflection model based on Knudsen's cosine law~\citep{ref27,ref28}. This law assumes that after the collision, the reflection direction follows a specific angular distribution: The probability of a reflection angle $\varepsilon$ (relative to the normal of the wall) is proportional to $\cos\varepsilon$. In the simulation implementation, a collision is identified when the atomic trajectory intersects the boundary of the cell, and the post-collision direction of the atom is randomly sampled according to this distribution. The speed after collision is denoted by \(v_2\). Since atoms with higher velocities are more likely to collide with the cell wall during their motion, a flux-weighted speed distribution is adopted so that the speed distribution of atoms after collisions matches the initial two-dimensional Maxwell-Boltzmann distribution, preserving the system's thermal equilibrium. The flux-weighted velocity distribution can be expressed as:
\begin{equation}
g(v) = \sqrt{\frac{2}{\pi}} \frac{v^2}{\sigma^3} \exp\left(-\frac{v^2}{2\sigma^2}\right).
\label{eq:v}
\end{equation}
In the Monte Carlo simulation, the atom-light coupling is not treated as a constant. Instead, for each atom, the instantaneous coupling strength is determined by the local probe intensity along its trajectory inside the Gaussian probe beam. As a result, the coupling $\kappa(t)$ fluctuates in time as the atom moves across regions with different optical intensities, which gives rise to transit-induced coupling inhomogeneity.

The decoherence of atomic spins arises primarily from various relaxation mechanisms, including atom-atom collisions, atom-wall collisions, and spontaneous emission. These effects lead to the decay of ground-state coherence, thereby shortening the coherence lifetime. In this simulation, two main decoherence mechanisms are considered: atom-wall collisions and spontaneous emission. As for the decoherence induced by atom-wall collisions, we adopt one of the classical approaches, namely, random phase resetting: when an atom collides with the rubidium cell wall, its spin state is randomly reset with a certain probability to simulate the random decoherence caused by the collision. For the decoherence induced by spontaneous emission, we use \(\gamma\) to denote the decay rate. However, if spontaneous emission is modeled merely as a classical exponential decay term, it would violate the fundamental commutation relations satisfied by the spin operators in quantum mechanics, leading to unphysical results. Therefore, a Langevin noise term~\citep{ref29} is introduced to maintain the completeness of quantum mechanics. The equations of motion for the atomic collective spin operator \(\hat{x}_A\) and \(\hat{p}_A\) can now be written as:
\begin{equation}
\begin{split}
\dot{\hat{x}}_A &= \kappa(t) \hat{p}_L^{in}(t) \cos(\Omega t) - \gamma \hat{x}_A(t) + \sqrt{2\gamma} \hat{f}_A(t), \\
\dot{\hat{p}}_A &= -\kappa(t) \hat{p}_L^{in}(t) \sin(\Omega t) - \gamma \hat{p}_A(t) + \sqrt{2\gamma} \hat{f}_B(t),
\end{split}
\label{eq:Langevin}
\end{equation}
where $\hat{f}_A$ and $\hat{f}_B$ are Langevin noise operators with appropriate variance. The decay rate \(\gamma\) includes contributions from probe-induced decoherence and other relaxation processes. In particular, the probe-induced part depends on the local probe intensity along the atomic trajectory. For different probe beam sizes at a fixed total optical power, the higher local intensity of a smaller beam is  automatically %AUTHOR: We have added 'automatically'.
compensated by the shorter residence time of the atoms inside the beam, so the spatially and temporally averaged spontaneous-emission-induced decoherence remains approximately unchanged. In the presence of Larmor precession, the measured spin component rotates in time, so the atom-light interaction is no longer a QND measurement of a fixed atomic quadrature $\hat{p}_A$. To recover an effective QND measurement, the atom-light coupling can be modulated in time; i.e., $\kappa(t)$ is modulated at %AUTHOR: We have changed 'with' to 'at'.
twice the Larmor frequency. In practice, this corresponds to a stroboscopic probing scheme, in which the same atomic quadrature is repeatedly measured \citep{ref7,ref6,ref30}. In this way, the measurement remains effectively QND despite the Larmor evolution.

\subsection{Degree of Spin Squeezing}

There are various definitions of spin squeezing parameters. In this paper, we adopt the spin squeezing parameter proposed by Wineland et al.~\citep{bu6,bu7}, which is given by
\begin{equation}
\xi^2 = \frac{(\Delta \phi)^2}{(\Delta \phi)_{\text{CSS}}^2},
\label{eq:wineland}
\end{equation}
which denotes the ratio between the phase resolution of the squeezed state and that of an ideal coherent spin state, which indicates the metrological gain of the entangled state.
%%%%%%%%%%%%%%%%%%%%%%%%%%%%

By extracting the atomic spin information from the light signal, the degree of spin squeezing can be computed via covariance and conditional variance analysis. The raw data are processed in three steps. The first step is performing a covariance analysis on the signal \( x_L^{out}\), which is represented as a matrix \(N \times n\), where \(n\) is the number of samplings, and \(N\) indicates the number of repeated measurements. After processing the covariance, %AUTHOR: We have changed 'covariance processing' to 'processing the covariance'.
an \(n \times n\) matrix is obtained. Its diagonal elements correspond to the variances in the measured signals at different times, while the off-diagonal elements represent the degree of correlation between signals at different times. Second, the conditional variance is calculated. The conditional spin squeezing in a QND measurement arises from the Faraday interaction between light and atoms, which entangles them and imprints atomic information onto the light signal. This constitutes a form of forward-predictive conditional spin squeezing. According to the definition of conditional variance, under the measurement result of \(x_L\), %AUTHOR: We have changed 'condition' to 'measurement result'.
the conditional variance of the atomic spin \(\hat{p}_A\) can be expressed as:
\begin{equation}
\mathrm{Var}(\hat{p}_A|\hat{x}_L)=\mathrm{Var}(\hat{p}_A)-
\frac{\mathrm{Cov}^2(\hat{p}_A,\hat{x}_L)}{\mathrm{Var}(\hat{x}_L)},
\label{eq:CondVar}
\end{equation}
where \(\operatorname{Cov}(\hat{p}_A, \hat{x}_L)\) is the correlation strength between \(\hat{x}_L\) and \(\hat{p}_A\).
Finally, the projection noise limit (PNL) needs to be calibrated. Experimentally, the PNL is calibrated at \linebreak 4/5 times~\citep{ref26} %AUTHOR: We have changed '0.8' to '4/5'.
the thermal state noise to evaluate the squeezing; theoretically, the stationary-atom noise can be used as the SQL reference.

%%%%%%%%%%%%%%%%%%%%%%%%%%%%%%%%%%%%%%%%%%
\section{Experimental Setup}

\begin{figure}[H]
%\centering
\includegraphics[width=0.95\textwidth]{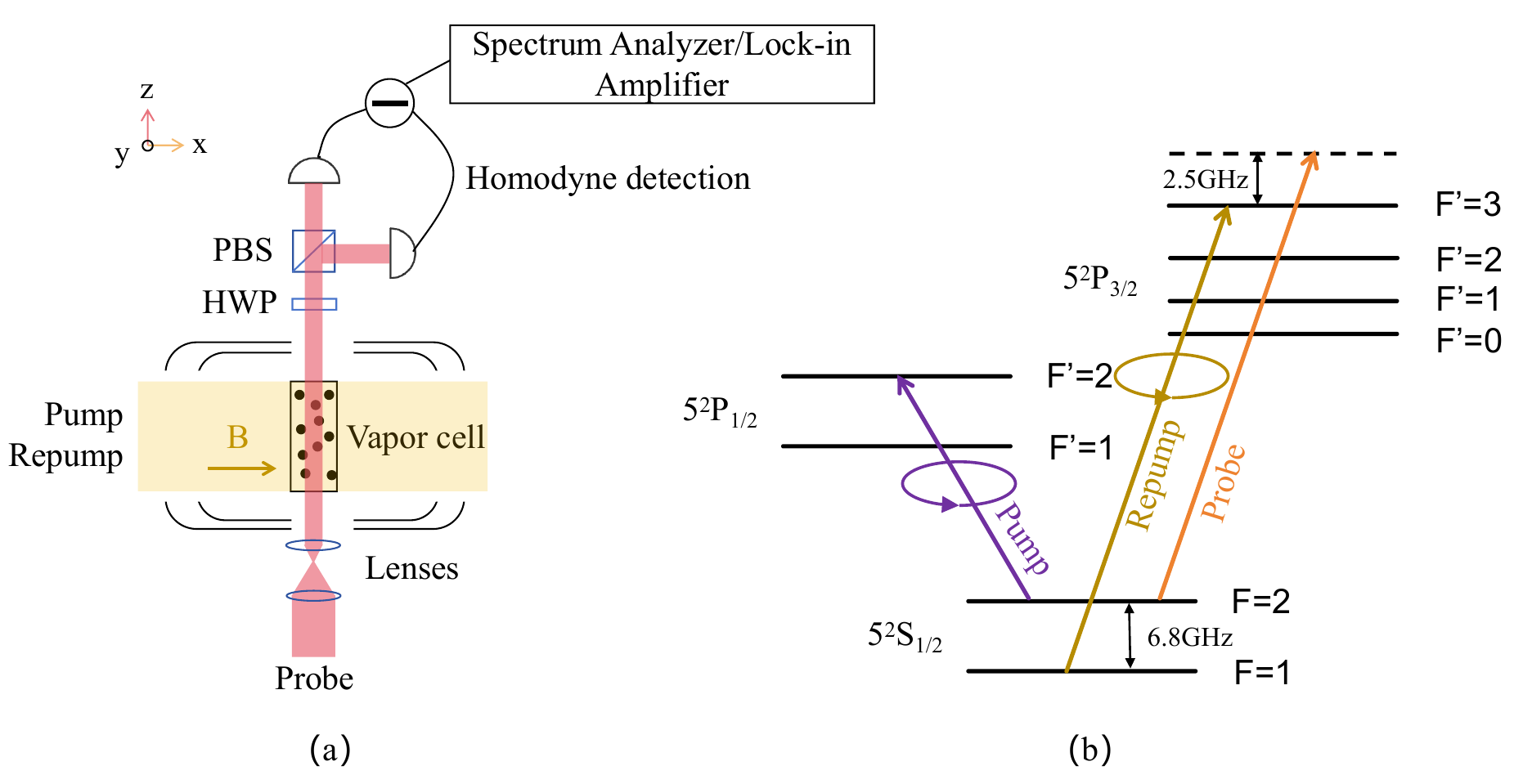}
\caption{(\textbf{a}) Experimental %MDPI: Please check if arrow, dashed line and symbols need add explanations.
%AUTHOR: We have added explanations where necessary. These have been highlighted in yellow in the figure caption.
 setup for quantum nondemolition measurement. PBS: polarizing beam splitter; HWP: half-wave plate. (\textbf{b}) Energy level diagram of the interaction between the light fields and $^{87}$Rb atoms. The arrows indicate the energy levels coupled by the pump, repump, and probe beams; the circles (attached to the arrows) denote circularly polarized light.} 
\label{fig:setup}
\end{figure}

The experimental setup and optical configuration for QND measurement are shown in Figure~\ref{fig:setup}a. A $^{87}$Rb vapor cell was positioned within a five-layer magnetic shield, which was used to shield against the geomagnetic field and external stray magnetic fields. A carefully designed coil assembly between the cell and the magnetic shield generated a bias magnetic field along the $x$-axis, whose amplitude %AUTHOR: We have changed 'strength' to 'amplitude'.
determined the Larmor precession frequency of the atomic spins. The cell, with dimensions of $3\text{ mm} \times 3\text{ mm} \times 20\text{ mm}$, was coated with an anti-relaxation coating and maintained at $58\,^\circ\mathrm{C}$ and contained approximately $10^{10}$ atoms. However, in our theoretical simulations, for simplicity, we used a cylindrical cell shape with equal cross-sectional areas, an approximation that proved to maintain good agreement with the experiment using rectangular cells~\citep{ref20}. This approximation was valid because the transit noise spectrum is governed primarily by the statistics of atoms crossing %AUTHOR: We have changed 'atomic crossings' to 'atoms crossing'.
through the probe beam and is therefore much more sensitive to the beam size than to the detailed shape of the cell cross-section. 

Three laser fields interacted with the \(^{87}\)Rb atoms: the pump and repump beams propagated along the \(x\)-axis, while a probe beam propagated along the \(z\)-axis. The corresponding energy %AUTHOR: We have removed 'coupling'.
level diagram is shown in Figure~\ref{fig:setup}b. The 795\,nm pump beam and the 780\,nm repump beam, both circularly polarized, coupled to the D\(_1\) and D\(_2\) lines, respectively, and optically pumped the atoms into the desired polarized state. A holding magnetic field of about 0.7 Gauss was applied along the polarized spin direction corresponding to a 500 kHz Larmor frequency, around which the spin noise was measured via a lock-in amplifier. This frequency was chosen as a compromise between increased technical noise at lower frequencies and magnetic field inhomogeneity at higher fields. The 780\,nm probe beam was linearly polarized, which coupled the \( ^{87}\text{Rb} \) atoms to the \( 5S_{1/2}F = 2 \rightarrow 5P_{3/2} \) transition. We locked the probe beam at a blue detuning of 2.5\,GHz using the dichroic atomic vapor laser lock (DAVLL) technique~\citep{bu11} and a fiber electro-optic modulator (EOM) for frequency shifting. In this experiment, the peak power of the probe beam was 5 mW. With a duty cycle of 0.1 in stroboscopic probing, the time-averaged power of the probe light was \linebreak 500 $\upmu$W. The probe beam with a Gaussian-profile passed through a lens group and then entered the \( ^{87}\text{Rb} \) vapor cell to interact with the atoms. The beam diameter was measured using a Thorlabs BP209-VIS/M beam profiler, which defined the beam diameter as the width at which the intensity drops to \(1/e^2\) of the peak value. In the theoretical model, the same definition for the beam diameter was adopted. After exiting the cell, the beam traveled through a half-wave plate and a polarizing beam splitter followed by a balanced detector to measure the Faraday rotation signal. The output signal was fed into either a spectrum analyzer or a lock-in amplifier for spectral analysis.

\section{Results}
\begin{figure}[H]
%\centering
\includegraphics[width=0.95\textwidth]{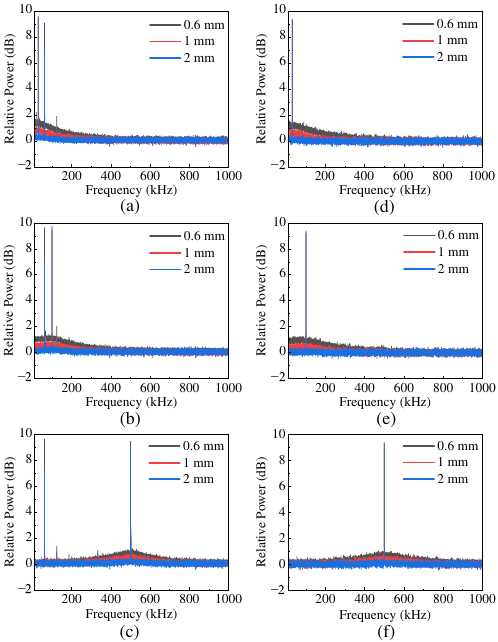}
\caption{Atomic spin %MDPI: Please note that changes to the position/size of figures or tables may occur during the production stage.
%AUTHOR: Confirmed. Thank you.
 noise spectra at different Larmor frequencies and probe beam sizes. Experiment (\textbf{left panels}) for Larmor frequencies of (\textbf{a}) 30~kHz; (\textbf{b}) 100~kHz; (\textbf{c}) 500~kHz. Monte Carlo simulation (\textbf{right panels}) for Larmor frequencies of (\textbf{d}) 30~kHz; (\textbf{e}) 100~kHz; (\textbf{f}) 500~kHz. }
\label{fig:spec}
\end{figure}

The atomic spin noise spectra corresponding to Larmor frequencies of 30~kHz, 100~kHz, and 500~kHz are shown in Figure~\ref{fig:spec}. The left panels represent the experimental spectra (averaged over 5000 measurements) measured with a spectrum analyzer, while the right panels show the corresponding Monte Carlo simulations (averaged over 2500 runs, which was a trade-off between SNR and computational load, as detailed in %AUTHOR: We have removed 'the'.
Appendix \ref{ave}). The experimental results are in good qualitative agreement with the theoretical predictions. However, in the measured spectra, in addition to the narrow resonance peaks at the Larmor frequency, several spurious peaks (e.g., at 62~kHz and 124~kHz) were observed. These arose from environmental electromagnetic interference coupled into the detection system, and the spurious noises %AUTHOR: We have changed 'coupling strength' to 'spurious noises'.
fluctuated over time (see Appendix \ref{PD} for a detailed discussion). All spectra exhibit a characteristic structure consisting of narrow peaks superimposed on a broad background. The narrow peaks correspond to the intrinsic atomic spin projection noise centered at the Larmor frequency, and the integrated absolute noise (area in linear scale) is contributed by all the atoms in the vapor cell, independent of the beam size~\cite{ref20}. Their linewidths are determined by the coherence lifetime of the atomic ground state. In contrast, the broad background originates from atomic transit noise, which arises from atoms moving into and out of the probe beam and experiencing position-dependent atom-light coupling. A smaller probe beam diameter leads to a more pronounced broad background, reflecting the enhanced contribution of atomic transit noise. This work focuses mainly on the influence of the broad background structure on spin squeezing.

\begin{figure}[H]
%\centering
\includegraphics[width=0.55\textwidth]{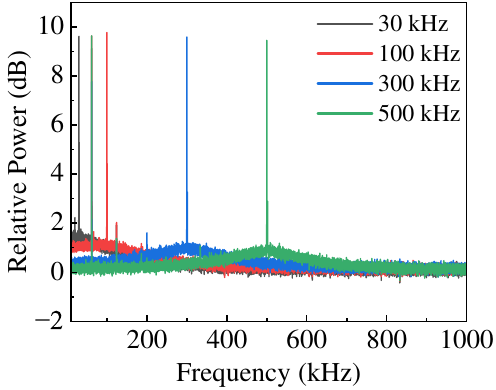}
\caption{ Variation in atomic transit noise with Larmor frequency in experiments.}
\label{fig:sq}
\end{figure}

As shown in Figure~\ref{fig:sq}, when the diameter of the beam is 0.6~mm, the visibility of this broad structure is strongly affected by the Larmor frequency. At relatively low Larmor frequencies, the width of the transit noise background is broad enough that the negative frequency component overlaps significantly with the positive frequency component in the detection band (a detailed explanation of this overlap is provided in Appendix \ref{pos}). %AUTHOR: We have removed 'the'.
This overlap causes the two components to add together, thereby elevating the observed broad background. As the Larmor frequency increases, the overlap is progressively reduced, and the broad background correspondingly becomes less pronounced.

In our spin squeezing experiment, the Larmor precession frequency was set to 500 kHz. At this frequency, the transit noise was slightly lower than that at lower Larmor frequencies. %AUTHOR: We have revised this sentence to improve clarity.
Experimentally, a lock-in amplifier selected a narrow detection window centered at this frequency, effectively suppressing out-of-band noise. In the theoretical model, we explicitly simulated this lock-in detection process by demodulating the probe signal and extracting the spectral component around 500~kHz, thereby ensuring a consistent comparison between theory and experiment.

\begin{figure}[H]
\includegraphics[width=0.95\textwidth]{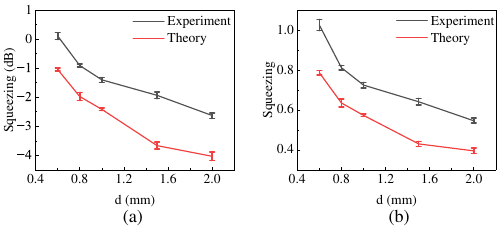}
\captionof{figure}{Calculated and
 measured spin squeezing at Larmor frequency of 500~kHz for different probe beam diameters. Both the experimental and theoretical values of the effective atom--light interaction strength \(\kappa\) are $1.61~\text{ms}^{-1/2}$, with total laser power fixed. Here, the squeezing was evaluated using a retrodictive estimator based on the past quantum state techniques, which improved the inference of the conditional atomic state from the measurement record. The corresponding effective \(\kappa^2T_2\) was 2.26 in the experiment, and the theoretical value differs from it by 0.22\%, where \(T_2\) is the transverse relaxation time. (\textbf{a}) Logarithmic scale; (\textbf{b}) linear scale.}
\label{fig:sq1}
\end{figure}

Squeezing parameters at 500~kHz for different probe beam diameters were further explored, and the results are shown in Figure~\ref{fig:sq1}. In our experiment, we employed a retrodiction technique~\cite{ref26} to increase spin squeezing, based on the past quantum state formalism introduced by K.~M{\o}lmer et~al.~\cite{PhysRevLett.111.160401} to improve the estimation of the conditional quantum state. In this formalism, the quantum state of a system at time $t$ depends not only on the measurement record obtained before $t$ but also on the information acquired after $t$, thereby enabling a more accurate estimate of the system state at time $t$. Each squeezing value plotted is the average of five independent experimental runs; the error bars correspond to the standard deviation of these runs. A difference of $2.7 \pm 0.2$ dB was observed between the \linebreak 0.6 mm and 2 mm beams experimentally, similar %AUTHOR: We have changed 'compared' to 'similar'.
to a theoretical prediction of \mbox{$3.0 \pm 0.3$ dB}. The 0.3 dB difference (between 2.7~dB and 3~dB) is at the level of the combined theoretical and experimental uncertainties.

\vspace{-10pt}

As the ratio between the probe beam radius and the vapor cell radius $R_p/R_c$ decreases, the squeezing performance increasingly degrades. When the beam is much smaller than the cell size, atoms experience stronger spatial variations in the light intensity during their motion, leading to intensified transit-induced fluctuations. Consequently, transit noise plays a more significant role in limiting the attainable squeezing, highlighting the importance of matching the probe beam size to the cell size. We note that when the beam size was 2~mm, there was a 1.4~dB discrepancy in spin squeezing between theory and experiment, which arose from experimental imperfections that were not fully captured in the model, including less-than-unity atomic polarization (0.22~dB), %AUTHOR: We have changed 'polarization imperfection' to 'less-than-unity atomic polarization'.
coherence loss (0.27~dB) and other technical noise such as electromagnetic noise.

Furthermore, the observed change %AUTHOR: We have changed 'difference' to 'change'.
in the degree of spin squeezing with beam sizes %AUTHOR: We have added 'with beam sizes'.
cannot be explained solely by the change %AUTHOR: We have changed 'increase' to 'change'.
in the transit-noise background. In our experiment, although the spin squeezing differs by about 2.7~dB between probe-beam sizes of 0.6~mm and 2~mm, the corresponding change in the transit-noise background at the Larmor frequency is significantly smaller (about 0.7~dB). This indicates that the degradation of squeezing is influenced by more than the net transit-noise level alone. In QND-based spin squeezing, another important contribution arises from the temporal decorrelation of the atomic spin projection signal induced by the transit process itself. As atoms move out of the probe beam, they cease to contribute to the measured signal, while atoms entering the beam can contribute new (different) spin projection signal. As a result, the temporal correlation of the QND readout is weakened, which worsens the conditional variance and further limits the achievable spin squeezing, especially for smaller probe beams.

\begin{figure}[H]
%\centering
\includegraphics[width=0.95\textwidth]{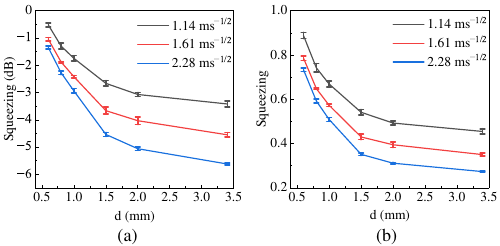}
\caption{Theoretical study %MDPI: Please change the hyphen (-) into a minus sign (−, “U+2212”) in the figure, e.g., “-1” should be “−1”.
%AUTHOR: We have changed the hyphens (-) to minus signs (−, "U+2212") in the figure as requested.
 of spin squeezing versus beam size under higher probe intensity conditions (labeled with effective atom-light interaction strength \(\kappa\), see text). At each \(\kappa\), total laser power is fixed for different beam diameters. (\textbf{a}) Logarithmic scale; (\textbf{b}) linear scale.}
\label{fig:hiI}
\end{figure}

In principle, a stronger atom-light interaction leads to a higher degree of spin squeezing, since more atomic spin information is imprinted onto the probe field during the QND measurement. However, in our experiment, increasing the probe beam power beyond the current value would have driven the photodiode out of its linear response region for photon-shot-noise-limited performance. Therefore, the behavior of transit noise at higher optical intensities was not characterized experimentally. Instead, we performed a theoretical analysis of this regime, the results of which are presented in Figure~\ref{fig:hiI}. The values $1.14~\text{ms}^{-1/2}$, $1.61~\text{ms}^{-1/2}$, and $2.28~\text{ms}^{-1/2}$ represent the effective atom-light interaction strength \(\kappa\). %AUTHOR: We have added '\(\kappa\)'.
In a stronger interaction regime, the requirement for spatially homogeneous coupling becomes increasingly stringent. For a Gaussian probe beam, atoms traversing different intensity regions experience substantially different interaction strengths, and the resulting transit-induced fluctuations are correspondingly amplified. As the coupling strength increases, these inhomogeneities can no longer be treated as a small perturbation but instead constitute the dominant factor limiting the attainable spin squeezing. Overall, while stronger probing is beneficial for enhancing the QND measurement strength, it also renders the experiment more susceptible to transit noise.

\vspace{-5pt}

\begin{figure}[H]
%\centering
\includegraphics[width=0.95\textwidth]{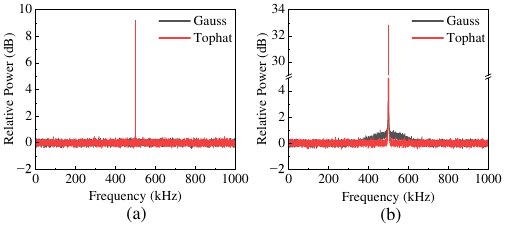}
\caption{Theoretical study of atomic spin noise spectra for Gaussian probe beam and tophat probe beam for (\textbf{a}) low coupling strength $\kappa = 1.61~\text{ms}^{-1/2}$; (\textbf{b}) high coupling strength $\kappa = 25.45~\text{ms}^{-1/2}$.}
\label{fig:gauss}
\end{figure}

To fully exploit the advantages of strong atom-light coupling, the interaction must be made more spatially uniform, for example, by expanding the probe beam relative to the atomic sample %AUTHOR: We have changed 'cloud' to 'sample'.
size or by using a beam profile with improved transverse homogeneity. However, we theoretically calculated the atomic transit noise spectra for a 3.4~mm tophat beam and Gaussian beam (same size of the vapor cell) at a Larmor frequency of 500~kHz. The results are shown in Figure~\ref{fig:gauss}. For a relatively small coupling strength ($\kappa = 1.61~\text{ms}^{-1/2}$) similar to that in our experiment, we did not observe an obvious difference between the two beam shapes in terms of transit noise contribution. In addition, we calculated the spin squeezing achievable with the tophat beam (beam size of 3.4~mm) as 4.75~dB and that with the Gaussian beam as 4.54~dB. Therefore, within the parameter regime in our experiment, neither beam shape exhibited a clear advantage over the other. For larger coupling strengths, the difference is slightly larger (Figure~\ref{fig:gauss}b). Overall, the difference between the two beam shapes is relatively small, which is due to the effect of motional averaging of the atoms in a coated cell. These results demonstrate that, in the strong coupling regime, suppressing transit noise via increasing the probe beam size is essential for achieving higher levels of spin squeezing.

\vspace{-5pt}

%%%%%%%%%%%%%%%%%%%%%%%%%%%%%%%%%%%%%%%%%%
\section{Conclusions}

In summary, we theoretically and experimentally investigated transit noise in a coated vapor cell and found good qualitative agreement between experiment and theory. We showed that the transit noise background %AUTHOR: We have corrected a spelling error.
becomes larger for smaller probe beams and has a stronger influence at lower Larmor frequencies under otherwise identical conditions. Moreover, its effect on spin squeezing is not determined solely by the transit noise background at the detection frequency but also by transit-induced decorrelation in the QND readout. Our results therefore suggest that, under the present experimental conditions, using a probe beam comparable to the cell cross-section is beneficial for suppressing transit-noise-induced degradation in spin squeezing. More broadly, this work provides insight into transit-related limitations in hot atom quantum sensors and may be relevant to other thermal atomic ensemble platforms, including optical magnetometers, electric field sensors, and atomic clocks.

\vspace{+6pt}
\authorcontributions{Conceptualization, Y.J., P.L., Y.X., Y.W. and J.D.; Methodology, Y.J., P.L. and J.D.; Validation, Y.J. and P.L.; Formal analysis, Y.J. and P.L.; Investigation, Y.J. and P.L.; Resources, Y.J.; Data curation, Y.J., P.L. and J.D.; Writing---original draft, Y.J. and P.L.; Writing---review \& editing, Y.X., Y.W. and J.D.; Visualization, Y.J.; Supervision, Y.X., Y.W. and J.D.; Project administration, Y.X. and J.D.; Funding acquisition, Y.X. and Y.W. All authors have read and agreed to the published version of the manuscript. %MDPI: We added the Author Contributions based on the information submitted online at susy.mdpi.com, and we changed the authors’ names into an abbreviated format. Please confirm.
%AUTHOR: We have checked the Author Contributions and the abbreviated author names. We confirm that they are correct as provided.
}

\funding{This %MDPI: Information regarding the funder and the funding number should be provided. Please check the accuracy of funding data and any other information carefully. No further modification is allowed after confirmation.
%AUTHOR: We have checked the funding information in the manuscript and confirm that it is accurate.
 research was funded by the Innovation Program for Quantum Science and Technology under grant No. 2023ZD0300900, National Natural Science Foundation of China (grant Nos. 12027806, 12161141018 and 62505164), Fund for Shanxi “1331 Project”, and talent introduction program of Sichuan University of Science \& Engineering (No. H40125120).}

\dataavailability{Data underlying the results presented in this paper are not publicly available at this time but may be obtained from the authors upon reasonable request.} 

\acknowledgments{We thank Mingyong Jing %MDPI: Titles (e.g., Dr., Mr., and Prof.) should NOT be used in the Acknowledgments section. We have removed them. Please confirm.
%AUTHOR: Confirmed. Thank you.
 for helpful discussions.}

\conflictsofinterest{The authors declare no conflicts of interest.}

\appendix
\section{\phantom{This is hidden}}
\label{kappa and p}
\numberwithin{equation}{section}
\setcounter{equation}{0} 
\setcounter{figure}{0}

%\makeatletter
%
%\setcounter{equation}{0}

%\@addtoreset{equation}{section}

\renewcommand{\theequation}{A\arabic{equation}}
%\makeatletter

The strength of the atom-light interaction $\kappa$ can be expressed as%MDPI: We revised the equation label in appendix according to our journal rule, please confirm.
%AUTHOR: Confirmed. Thank you.
~\citep{bu12,ref7,ref25}:
\begin{equation}
\kappa = -\frac{\Gamma \lambda^2}{16\pi A \Delta} a_1 \sqrt{ \frac{P N_A}{\hbar \omega} },
\label{eq:k}
\end{equation}
where $\Gamma = 2\pi \times 6.07 \, \mathrm{MHz}$ is the nature width of the excited state, %AUTHOR: We have revised the definition of \(\Gamma\).
$\lambda = 780 \, \mathrm{nm}$ is the probe wavelength, $A$ is the atom-light interaction area, $P$ is the optical power and $\Delta = -2\pi \times 2.5\ \text{GHz}$ is the detuning of the probe. $a_1$ is the vector interaction coefficient~\citep{bu12} and can be expressed as:
\begin{equation}
a_1 = \frac{\sqrt{2}}{100} \left( -\frac{15}{1 - \Delta_{13}/\Delta} - \frac{25}{1 - \Delta_{23}/\Delta} + 140 \right).
\label{eq:a1}
\end{equation}
Here, $\Delta_{13} = 2\pi \times 423.60 \, \text{MHz}$ and $\Delta_{23} = 2\pi \times 266.65 \, \text{MHz}$ are the energy level differences between $F' = 1$ and $F' = 3$ and between $F' = 2$ and $F' = 3$ in the hyperfine structure of the excited state $5^2\text{P}_{3/2}$, respectively. Equation~\eqref{eq:k} gives the effective coupling strength in the homogeneous coupling approximation. In the Monte Carlo model, the instantaneous coupling $\kappa(t)$ is obtained by replacing the uniform probe intensity with the local intensity sampled along the atomic trajectory.

\section{\phantom{This is hidden}}
\label{ave}
\numberwithin{equation}{section}
\setcounter{equation}{0} 
\setcounter{figure}{0}
\renewcommand{\thefigure}{A\arabic{figure}}

\vspace{+5pt}

\begin{figure}[H]
%\centering
\includegraphics[width=0.95\linewidth]{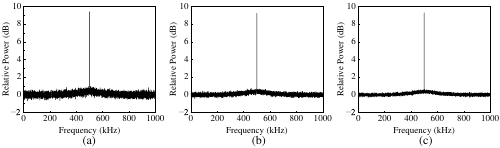} 
\caption{Calculated 
 atomic spin noise spectra at different averaging numbers for a 1~mm beam at 500~kHz Larmor frequency. Monte Carlo simulation for averaging numbers of (\textbf{a}) 1000; (\textbf{b}) 2500; \linebreak (\textbf{c}) 5000. }
\label{fig:avg_comparison}
\end{figure}

Figure~\ref{fig:avg_comparison} %MDPI: We revised the figure name in appendix, please confirm. Same as below.
%AUTHOR: Confirmed. Thank you.
shows the simulated noise spectra for a 1~mm beam at a 500~kHz Larmor frequency, obtained with 1000, 2500 and 5000 averages. As the figure shows, increasing the number of averages improves the signal-to-noise ratio (SNR), but does not change the overall spectral structure. However, running the simulation with 5000 averages takes approximately twice the time of 2500 averages, making it computationally heavy. Therefore, considering that the improvement from 2500 to 5000 averages is marginal, we chose 2500 averaging times. 

\vspace{-5pt}

\section{\phantom{This is hidden}}
\label{PD}
\numberwithin{equation}{section}
\setcounter{equation}{0} 
\setcounter{figure}{1}
\renewcommand{\thefigure}{A\arabic{figure}}

\vspace{+5pt}

\begin{figure}[H]
%\centering
\includegraphics[width=0.55\linewidth]{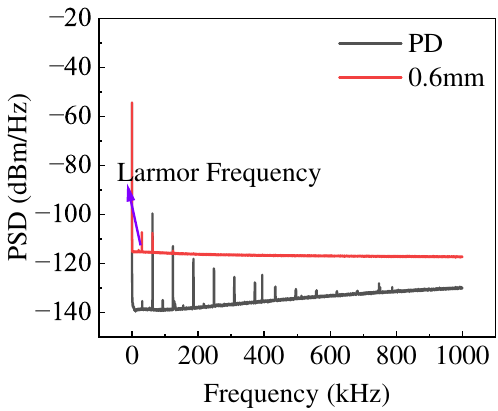} 
\caption{Background noise of the %AUTHOR: We have added 'the'.
photodetector (black) and measured noise spectrum for 0.6~mm beam at 30~kHz Larmor frequency (red). PSD: power spectral density.}
\label{fig:emi_spectra}
\end{figure}

Figure~\ref{fig:emi_spectra} shows two spectra: the background noise of the photodetector (black curve) and the measured noise spectrum for a 0.6~mm beam at a 30~kHz Larmor frequency (red curve, taken from Figure~\ref{fig:spec}). In the measured spectrum, distinct peaks at 62~kHz and 124~kHz are clearly visible, and these noise peaks rise above the atomic spin noise. These peaks originate from environmental electromagnetic interference (EMI) that couples into the detection system. The EMI strength varies over time and cannot be removed by simply subtracting the background light noise. However, the data used to extract the degree of spin squeezing are demodulated at the Larmor frequency, which was chosen away from the EMI frequency.

\vspace{-5pt}

\section{\phantom{This is hidden}}
\label{pos}
\numberwithin{equation}{section}
\setcounter{equation}{0} 
\setcounter{figure}{2}
\renewcommand{\thefigure}{A\arabic{figure}}

\vspace{+5pt}

\begin{figure}[H]
%\centering
\includegraphics[width=0.55\linewidth]{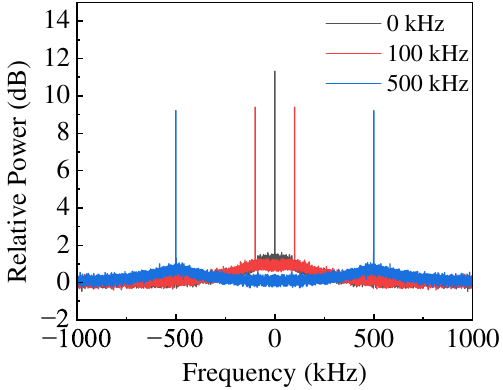}
\caption{Calculated atomic noise spectra for a 0.6~mm beam at Larmor frequencies of 0~kHz, 100~kHz, and 500~kHz.}
\label{fig:overlap}
\end{figure}

The positive-negative frequency overlap effect at low Larmor frequencies $\Omega$ is illustrated in Figure~\ref{fig:overlap}, which shows the theoretically calculated atomic noise spectra for a 0.6~mm beam at Larmor frequencies of 0~kHz (black), 100~kHz (red), and 500~kHz (blue). The Larmor precession of the atomic spin produces an oscillatory signal at frequency $\Omega$, whose Fourier spectrum ideally contains two symmetric sidebands at $+\Omega$ and $-\Omega$. However, in actual measurements, we recorded only a single sideband by using heterodyne detection and filtering. When the Larmor frequency is lower than the width of the broad structure in the noise spectrum contributed by the transit process, the positive and negative sidebands overlap, leading to an apparent increase in the measured noise level. This effect is clearly shown by the red curve in the figure.

%\vspace{-5pt}

\begin{adjustwidth}{-\extralength}{0cm}
%\centering %% If there is a figure in wide page, please release command \centering
%%%%%%%%%%%%%%%%%%%%%%%%%%%%%%%%%%%%%%%%%%

\reftitle{References}

\PublishersNote{}

\end{adjustwidth}

\begin{thebibliography}{999}

\bibitem[Zhan and Xie(2020)]{bu1}
Zhan, M.; Xie, X.
\newblock Precision measurement physics: Physics that precision matters.
\newblock {\em Natl. Sci. Rev.} {\bf 2020}, {\em 7},~1795.

\bibitem[Wang et~al.(2023)Wang, Huang, Guo, Jiang, Kang, Su, Qin, Ji, Hu, Peng,
atom--lightand Budker]{bu2}
Wang, Y.; Huang, Y.; Guo, C.; Jiang, M.; Kang, X.; Su, H.; Qin, Y.; Ji, W.; Hu,
 D.; Peng, X.; et~al.
\newblock Search for exotic parity-violation interactions with quantum spin
 amplifiers.
\newblock {\em Sci. Adv.} {\bf 2023}, {\em 9},~eade0353.

\bibitem[d’Armagnac~de Castanet et~al.(2024)d’Armagnac~de Castanet,
 Des~Cognets, Arguel, Templier, Jarlaud, M{\'e}noret, Desruelle, Bouyer, and
 Battelier]{bu3}
d’Armagnac~de Castanet, Q.; Des~Cognets, C.; Arguel, R.; Templier, S.;
 Jarlaud, V.; M{\'e}noret, V.; Desruelle, B.; Bouyer, P.; Battelier, B.
\newblock Atom interferometry at arbitrary orientations and rotation rates.
\newblock {\em Nat. Commun.} {\bf 2024}, {\em 15},~6406.

\bibitem[Li et~al.(2025)Li, Chen, Yang, Xu, Huang, Luo, Zhao, Niu, Liu, Yao,
 et~al.]{bu4}
Li, C.-Y.; Chen, L.-K.; Yang, X.; Xu, Z.-Y.; Huang, M.-Q.; Luo, Y.; Zhao, Y.-H.;
 Niu, X.-W.; Liu, Z.-W.; Yao, H.-J.; %MDPI: We revised the author names format, please confirm. Same as below.
 %AUTHOR: Confirmed. Thank you.
 et~al.
\newblock Drift-free continuous gravity measurement and application analysis of
 a high-precision atom gravimeter.
\newblock {\em Phys. Rev. Appl.} {\bf 2025}, {\em 24},~014045.

\bibitem[Hinkley et~al.(2013)Hinkley, Sherman, Phillips, Schioppo, Lemke,
 Beloy, Pizzocaro, Oates, and Ludlow]{ref1}
Hinkley, N.; Sherman, J.A.; Phillips, N.B.; Schioppo, M.; Lemke, N.D.; Beloy,
 K.; Pizzocaro, M.; Oates, C.W.; Ludlow, A.D.
\newblock An atomic clock with $10^{-18}$ instability.
\newblock {\em Science} {\bf 2013}, {\em 341},~1215--1218.

\bibitem[Bloom et~al.(2014)Bloom, Nicholson, Williams, Campbell, Bishof, Zhang,
 Zhang, Bromley, and Ye]{ref2}
Bloom, B.J.; Nicholson, T.L.; Williams, J.R.; Campbell, S.; Bishof, M.; Zhang,
 X.; Zhang, W.; Bromley, S.; Ye, J.
\newblock An optical lattice clock with accuracy and stability at the
 $10^{-18}$ level.
\newblock {\em Nature} {\bf 2014}, {\em 506},~71--75.

\bibitem[Facon et~al.(2016)Facon, Dietsche, Grosso, Haroche, Raimond, Brune,
 and Gleyzes]{ref3}
Facon, A.; Dietsche, E.K.; Grosso, D.; Haroche, S.; Raimond, J.M.; Brune, M.;
 Gleyzes, S.
\newblock A sensitive electrometer based on a Rydberg atom in a
 Schr{\"o}dinger-cat state.
\newblock {\em Nature} {\bf 2016}, {\em 535},~262--265.

\bibitem[Kominis et~al.(2003)Kominis, Kornack, Allred, and Romalis]{ref4}
Kominis, I.K.; Kornack, T.W.; Allred, J.C.; Romalis, M.V.
\newblock A subfemtotesla multichannel atomic magnetometer.
\newblock {\em Nature} {\bf 2003}, {\em 422},~596--599.

\bibitem[Pezz\`e et~al.(2018)Pezz\`e, Smerzi, Oberthaler, Schmied, and
 Treutlein]{ref5}
Pezz\`e, L.; Smerzi, A.; Oberthaler, M.K.; Schmied, R.; Treutlein, P.
\newblock Quantum metrology with nonclassical states of atomic ensembles.
\newblock {\em Rev. Mod. Phys.} {\bf 2018}, {\em 90},~035005.

\bibitem[Vasilakis et~al.(2015)Vasilakis, Shen, Jensen, Balabas, Salart, Chen,
 and Polzik]{ref6}
Vasilakis, G.; Shen, H.; Jensen, K.; Balabas, M.; Salart, D.; Chen, B.; Polzik,
 E.S.
\newblock Generation of a squeezed state of an oscillator by stroboscopic
 back-action-evading measurement.
\newblock {\em Nat. Phys.} {\bf 2015}, {\em 11},~389--392.

\bibitem[Bao et~al.(2020)Bao, Duan, Jin, Lu, Li, Qu, Wang, Novikova, Mikhailov,
 Zhao, et~al.]{ref7}
Bao, H.; Duan, J.; Jin, S.; Lu, X.; Li, P.; Qu, W.; Wang, M.; Novikova, I.;
 Mikhailov, E.E.; Zhao, K.F.; et~al.
\newblock Spin squeezing of \linebreak $10^{11}$ atoms by prediction and retrodiction
 measurements.
\newblock {\em Nature} {\bf 2020}, {\em 581},~159--163.

\bibitem[Hosten et~al.(2016)Hosten, Engelsen, Krishnakumar, and Kasevich]{ref8}
Hosten, O.; Engelsen, N.J.; Krishnakumar, R.; Kasevich, M.A.
\newblock Measurement noise 100 times lower than the quantum-projection limit
 using entangled atoms.
\newblock {\em Nature} {\bf 2016}, {\em 529},~505--508.

\bibitem[Ludlow et~al.(2015)Ludlow, Boyd, Ye, Peik, and Schmidt]{ref9}
Ludlow, A.D.; Boyd, M.M.; Ye, J.; Peik, E.; Schmidt, P.O.
\newblock Optical atomic clocks.
\newblock {\em Rev. Mod. Phys.} {\bf 2015}, {\em 87},~637--701.

\bibitem[Robinson et~al.(2024)Robinson, Miklos, Tso, Kennedy, Bothwell, Kedar,
 Thompson, and Ye]{ref10}
Robinson, J.M.; Miklos, M.; Tso, Y.M.; Kennedy, C.J.; Bothwell, T.; Kedar, D.;
 Thompson, J.K.; Ye, J.
\newblock Direct comparison of two spin-squeezed optical clock ensembles at the
 $10^{-17}$ level.
\newblock {\em Nat. Phys.} {\bf 2024}, {\em 20},~208--213.

\bibitem[Malia et~al.(2022)Malia, Wu, Mart{\'\i}nez-Rinc{\'o}n, and
 Kasevich]{ref11}
Malia, B.K.; Wu, Y.; Mart{\'\i}nez-Rinc{\'o}n, J.; Kasevich, M.A.
\newblock Distributed quantum sensing with mode-entangled spin-squeezed atomic
 states.
\newblock {\em Nature} {\bf 2022}, {\em 612},~661--665.

\bibitem[Allred et~al.(2002)Allred, Lyman, Kornack, and Romalis]{ref12}
Allred, J.; Lyman, R.; Kornack, T.; Romalis, M.V.
\newblock High-sensitivity atomic magnetometer unaffected by spin-exchange
 relaxation.
\newblock {\em Phys. Rev. Lett.} {\bf 2002}, {\em 89},~130801.

\bibitem[Martin~Ciurana et~al.(2017)Martin~Ciurana, Colangelo, Slodi{\v{c}}ka,
 Sewell, and Mitchell]{ref13}
Martin~Ciurana, F.; Colangelo, G.; Slodi{\v{c}}ka, L.; Sewell, R.; Mitchell, M.
\newblock Entanglement-enhanced radio-frequency field detection and waveform
 sensing.
\newblock {\em Phys. Rev. Lett.} {\bf 2017}, {\em 119},~043603.

\bibitem[Koschorreck et~al.(2010)Koschorreck, Napolitano, Dubost, and
 Mitchell]{ref14}
Koschorreck, M.; Napolitano, M.; Dubost, B.; Mitchell, M.
\newblock Sub-projection-noise sensitivity in broadband atomic magnetometry.
\newblock {\em Phys. Rev. Lett.} {\bf 2010}, {\em 104},~093602.

\bibitem[Leroux et~al.(2010)Leroux, Schleier-Smith, and Vuleti{\'c}]{ref15}
Leroux, I.D.; Schleier-Smith, M.H.; Vuleti{\'c}, V.
\newblock Implementation of cavity squeezing of a collective atomic spin.
\newblock {\em Phys. Rev. Lett.} {\bf 2010}, {\em 104},~073602.

\bibitem[Luo et~al.(2017)Luo, Zou, Wu, Liu, Han, Tey, and You]{ref16}
Luo, X.-Y.; Zou, Y.-Q.; Wu, L.-N.; Liu, Q.; Han, M.-F.; Tey, M.K.; You, L.
\newblock Deterministic entanglement generation from driving through quantum
 phase transitions.
\newblock {\em Science} {\bf 2017}, {\em 355},~620--623.

\bibitem[Franke et~al.(2023)Franke, Muleady, Kaubruegger, Kranzl, Blatt, Rey,
 Joshi, and Roos]{ref17}
Franke, J.; Muleady, S.R.; Kaubruegger, R.; Kranzl, F.; Blatt, R.; Rey, A.M.;
 Joshi, M.K.; Roos, C.F.
\newblock Quantum-enhanced sensing on optical transitions through finite-range
 interactions.
\newblock {\em Nature} {\bf 2023}, {\em 621},~740--745.

\bibitem[Jin et~al.(2024)Jin, Duan, Zhang, Zhang, Bao, Shen, Xiao, Jia, Wang,
 and Xiao]{ref18}
Jin, S.; Duan, J.; Zhang, Y.; Zhang, X.; Bao, H.; Shen, H.; Xiao, L.; Jia, S.;
 Wang, M.; Xiao, Y.
\newblock Concurrent spin squeezing and light squeezing in an atomic ensemble.
\newblock {\em Phys. Rev. Lett.} {\bf 2024}, {\em 133},~173604.

\bibitem[Hines et~al.(2023)Hines, Rajagopal, Moreau, Wahrman, Lewis,
 Markovi{\'c}, and Schleier-Smith]{ref19}
Hines, J.A.; Rajagopal, S.V.; Moreau, G.L.; Wahrman, M.D.; Lewis, N.A.;
 Markovi{\'c}, O.; Schleier-Smith, M.
\newblock Spin squeezing by Rydberg dressing in an array of atomic ensembles.
\newblock {\em Phys. Rev. Lett.} {\bf 2023}, {\em 131},~063401.

\bibitem[Duan et~al.(2025)Duan, Hu, Lu, Xiao, Jia, M{\o}lmer, and Xiao]{ref30}
Duan, J.; Hu, Z.; Lu, X.; Xiao, L.; Jia, S.; M{\o}lmer, K.; Xiao, Y.
\newblock Concurrent spin squeezing and field tracking with machine learning.
\newblock {\em Nat. Phys.} {\bf 2025}, {\em 21},~909--915.

\bibitem[Tang et~al.(2020)Tang, Wen, Cai, and Zhao]{ref20}
Tang, Y.; Wen, Y.; Cai, L.; Zhao, K.
\newblock Spin-noise spectrum of hot vapor atoms in an anti-relaxation-coated
 cell.
\newblock {\em Phys. Rev. A} {\bf 2020}, {\em 101},~013821.

\bibitem[Wang et~al.(2023)Wang, Jing, Zhang, Yuan, Zhang, Zhang, Xiao, and
 Jia]{ref21}
Wang, Z.; Jing, M.; Zhang, P.; Yuan, S.; Zhang, H.; Zhang, L.; Xiao, L.; Jia,
 S.
\newblock Noise analysis of the atomic superheterodyne receiver based on
 flat-top laser beams.
\newblock {\em Opt. Express} {\bf 2023}, {\em 31},~19909--19917.

\bibitem[Yang et~al.(2024)Yang, Zeng, Li, Tian, Ji, Hu, Song, and Jiao]{ref22}
Yang, X.; Zeng, G.; Li, J.; Tian, Z.; Ji, Y.; Hu, J.; Song, X.;
 Jiao, Y.
 Investigation of Atomic Transit Noise in Rydberg Receiver.
\newblock {\em J. Quantum Opt.} {\bf 2024}, {\em 30},~92--98. %MDPI: please check if the page number is correct.
%AUTHOR: Thank you for your reminder. We have checked and corrected the page numbers.
\newblock (In Chinese)

\bibitem[B{\ae}rentsen et~al.(2024)B{\ae}rentsen, Fedorov, {\O}stfeldt,
 Balabas, Zeuthen, and Polzik]{ref23}
B{\ae}rentsen, C.; Fedorov, S.A.; {\O}stfeldt, C.; Balabas, M.V.; Zeuthen, E.;
 Polzik, E.S.
\newblock Squeezed light from an oscillator measured at the rate of
 oscillation.
\newblock {\em Nat. Commun.} {\bf 2024}, {\em 15},~4146.

\bibitem[Hammerer et~al.(2010)Hammerer, S{\o}rensen, and Polzik]{bu6}
Hammerer, K.; S{\o}rensen, A.S.; Polzik, E.S.
\newblock Quantum interface between light and atomic ensembles.
\newblock {\em Rev. Mod. Phys.} {\bf 2010}, {\em 82},~1041--1093.

\bibitem[Holstein and Primakoff(1940)]{ref24}
Holstein, T.; Primakoff, H.
\newblock Field dependence of the intrinsic domain magnetization of a
 ferromagnet.
\newblock {\em Phys. Rev.} {\bf 1940}, {\em 58},~1098--1113.

\bibitem[Julsgaard(2003)]{ref25}
Julsgaard, B.
\newblock Entanglement and Quantum Interactions with Macroscopic Gas Samples.
\newblock Ph.D. Thesis, University of Aarhus, Aarhus, %MDPI: we removed the extra ``Denmark'', please confirm.
%AUTHOR: Confirmed. Thank you.
 Denmark, 2003.

\bibitem[Jin(2023)]{bu12}
Jin, S.
\newblock Joint Control of Spin Squeezing and Light Squeezing in a Hot Atomic
 Ensemble.
\newblock Ph.D. Thesis, Fudan University, Shanghai, China, 2023.
\newblock (In Chinese)

\bibitem[Bao(2019)]{ref26}
Bao, H.
\newblock Experimental Study of Spin Squeezing in Large Scale Coated Rubidium
 Vapor Cell.
\newblock Ph.D. Thesis, Fudan University, Shanghai, China, 2019.
\newblock (In Chinese)

\bibitem[Bohnet et~al.(2014)Bohnet, Cox, Norcia, Weiner, Chen, and
 Thompson]{bu8}
Bohnet, J.G.; Cox, K.C.; Norcia, M.A.; Weiner, J.M.; Chen, Z.; Thompson, J.K.
\newblock Reduced spin measurement back-action for a phase sensitivity ten
 times beyond the standard quantum limit.
\newblock {\em Nat. Photon.} {\bf 2014}, {\em 8},~731--736.

\bibitem[Sewell et~al.(2012)Sewell, Koschorreck, Napolitano, Dubost, Behbood,
 and Mitchell]{bu9}
Sewell, R.J.; Koschorreck, M.; Napolitano, M.; Dubost, B.; Behbood, N.;
 Mitchell, M.W.
\newblock Magnetic sensitivity beyond the projection noise limit by spin
 squeezing.
\newblock {\em Phys. Rev. Lett.} {\bf 2012}, {\em 109},~253605.

\bibitem[Schleier-Smith et~al.(2010)Schleier-Smith, Leroux, and
 Vuleti{\'c}]{bu10}
Schleier-Smith, M.H.; Leroux, I.D.; Vuleti{\'c}, V.
\newblock States of an ensemble of two-level atoms with reduced quantum
 uncertainty.
\newblock {\em Phys. Rev. Lett.} {\bf 2010}, {\em 104},~073604.

\bibitem[Xu et~al.(2013)Xu, Qu, Gao, Hu, and Xiao]{ref27}
Xu, Z.-X.; Qu, W.-Z.; Gao, R.; Hu, X.-H.; Xiao, Y.-H.
\newblock Linewidth of electromagnetically induced transparency under motional
 averaging in a coated vapor cell.
\newblock {\em Chin. Phys. B} {\bf 2013}, {\em 22},~033202.

\bibitem[Steckelmacher(1986)]{ref28}
Steckelmacher, W.
\newblock Knudsen flow 75 years on: The current state of the art for flow of
 rarefied gases in tubes and systems.
\newblock {\em Rep. Prog. Phys.} {\bf 1986}, {\em 49},~1083--1107.

\bibitem[Krauter(2011)]{ref29}
Krauter, H.
\newblock Generation and Application of Entanglement of Room Temperature
 Ensembles of Atoms.
\newblock Ph.D. Thesis, University of Copenhagen, Copenhagen, Denmark, 2011.

\bibitem[Wineland et~al.(1994)Wineland, Bollinger, Itano, and Heinzen]{bu7}
Wineland, D.J.; Bollinger, J.J.; Itano, W.M.; Heinzen, D.J.
\newblock Squeezed atomic states and projection noise in spectroscopy.
\newblock {\em Phys. Rev. A} {\bf 1994}, {\em 50},~67--88.

\bibitem[Marchant et~al.(2010)Marchant, H{\"a}ndel, Wiles, Hopkins, Adams, and
 Cornish]{bu11}
Marchant, A.L.; H{\"a}ndel, S.; Wiles, T.P.; Hopkins, S.A.; Adams, C.S.;
 Cornish, S.L.
\newblock Off-resonance laser frequency stabilization using the Faraday effect.
\newblock {\em Opt. Lett.} {\bf 2010}, {\em 36},~64--66.

\bibitem[Gammelmark et~al.(2013)Gammelmark, Julsgaard, and
 M\o{}lmer]{PhysRevLett.111.160401}
Gammelmark, S.; Julsgaard, B.; M\o{}lmer, K.
\newblock Past Quantum States of a Monitored System.
\newblock {\em Phys. Rev. Lett.} {\bf 2013}, {\em 111},~160401.

\end{thebibliography}
\end{document}